\documentclass{ewic-latex}

\usepackage{array,multirow,graphicx}
\usepackage{enumitem} 
\usepackage{float, stfloats}

\usepackage[pscoord]{eso-pic}

\newcommand{\placetextbox}[3]{
  \setbox0=\hbox{#3}
  \AddToShipoutPictureFG*{
    \put(\LenToUnit{#1\paperwidth},\LenToUnit{#2\paperheight}){\vtop{{\null}\makebox[0pt][c]{#3}}}%
  }%
}%

\begin{document}

\runningheads{Rhodes $\bullet$ Woolley}{Back to the Future Museum – Speculative Design for Virtual, Citizen-Curated Museums}

\conference{PREPRINT - Submitted to 38th International BCS Human-Computer Interaction Conference 2025}

\title{Back to the Future Museum – Speculative Design for Virtual Citizen-Curated Museums}

\authorone{
    Richard Rhodes\\
    School of Computer Science and Mathematics\\
    Keele University ST5 5BG\\
    ORCID: 0000-0002-6955-7216\\
    \email{r.t.rhodes@keele.ac.uk}
}

\authortwo{
    Sandra Woolley\\
    School of Computer Science and Mathematics\\
    Keele University ST5 5BG\\
    ORCID: 0000-0002-7623-2866\\
    \email{s.i.woolley@keele.ac.uk}
}

\begin{abstract}
This forward-looking paper uses speculative design fiction to explore future museum scenarios where citizen curators design and share immersive virtual reality museums populated with tangible heritage artefacts, intangible virtual elements and interactive experiences. The work also explores takeaway ‘asset packs’ containing 3D artefact models, curation assets, and interactive experiences, and we envisage a visit to the future museum, where the physical and virtual experiences interplay. Finally, the paper considers the implications of this future museum in terms of resources and the potential impacts on traditional museums. 
\end{abstract}

\maketitle
\begin{figure*}[bp ]
    \centering
    \includegraphics[width=1\linewidth]{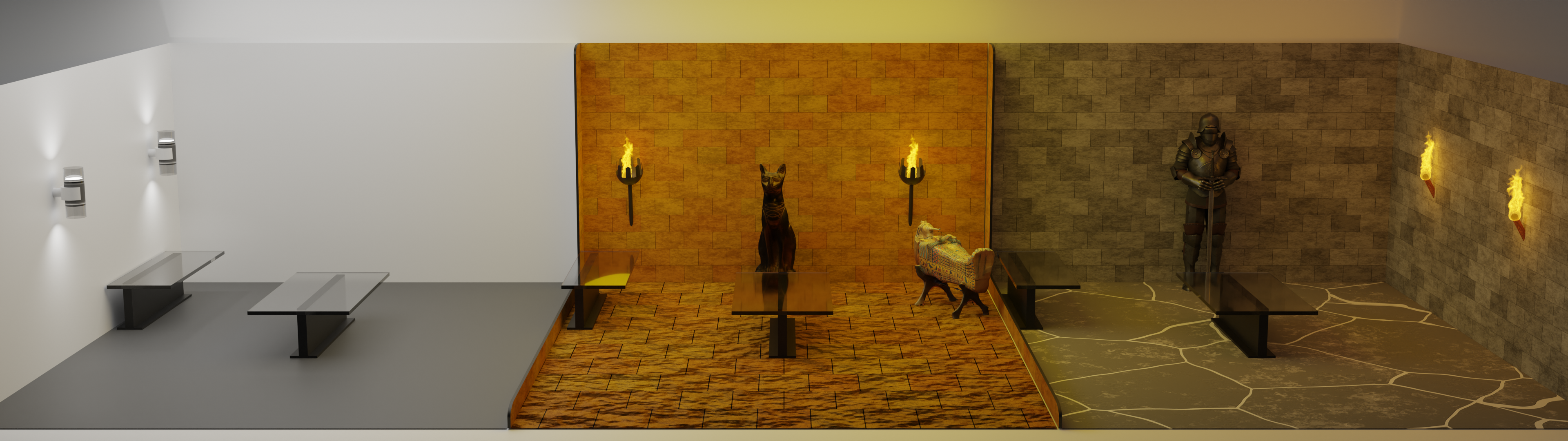}
    \caption{Alternative virtual galleries with style-appropriate curation assets for future museum citizen curators}
\end{figure*}

\placetextbox{0.6}{0.9}{\huge\color{red} \textbf{PREPRINT - Submitted to BCS HCI 2025}}

\section{Introduction}
Digital accompaniments to museum visits are not new; for example, apps supplementing museum experiences have been available for some time.  Gomes Barbosa, De Saboya, and Vaz Bevilaqua (2021) surveyed and evaluated 54 science museum apps. Noting variable app quality, the authors recommended guidance for app ease-of-use, functionality, quality graphics, structuring of information, and enabling users to label preferred exhibits.

Virtual Reality (VR) and Augmented Reality (AR) have also been used to support museum learning, producing favourable effects in both academic achievement and learner perceptions (Zhou, Chen, and Wang, 2022). Accessible and user-friendly AR in particular has been identified for its potential to offer museum visitors a unique experience (Artikova and Artikov, 2024). VR adoption, however, has been  constrained. Hall, et. al. (2019) explained that parental resistance is a significant barrier to the adoption of VR headsets for home use, observing that parents see Immersive Virtual Reality as a medium that isolates offspring from the family. These negative perceptions of VR could persist outside the home, and perhaps into museums and, indeed, museum professionals have expressed concerns around the logistics and engagement via VR in museums (Shehade and Stylianou-Lambert, 2020).

As illustrated in Figure 1, in this paper, we return to future heritage visions explored by Woolley, et. al. (2021) and Rhodes et al. (2024), and explore design fiction for future museums, envisioning both AR and VR for virtual curation spaces.

\section{Background}
After a period of closures during the coronavirus pandemic, museums reopened with a little more emphasis on open and adaptable digital tools than on guides, and handbooks (Parry and Dziekan, 2021). However, there is a significant divide between current physical (and often impressive) museums and (fragmentary) virtual alternatives. Ideally, there would be a virtual platform which enables collecting artefacts, curating them into personal galleries, and sharing them with others. Zidianakis et al. (2021) envisaged a virtual ‘invisible museum’ of digitized artefacts and virtual user exhibitions. 

Speculative design and design fiction have many applications in human-computer interaction. Flint, et. al. (2024) described design fictions as \textit{“narratives that occupy a time frame somewhere between the near present and a possible future. By taking possible or existing technology and extrapolating into their potential mundane use, we can make inferences about their benefits or pitfalls.”} We use these techniques to extend beyond Zidianakis et al.'s 'invisible museum', connecting both physical and virtual museums, and connecting and compiling ‘asset packs’ of artefacts, curation assets, and experiences to create richly populated and customised galleries.

Our research case study is Ancient Mesopotamian history. Mesopotamia, from the Greek \textit{“between two rivers”}, was the ‘cradle of civilisation’ some 5,000 years ago nestled between the Euphrates and Tigris rivers. Ancient Mesopotamians are credited with many innovations, including humankind’s first mathematics, first writing, and even the first customer complaint, as well as the first ‘Royal Game of Ur’ board game, of which there is a VR implementation (Pietroszek, Agraraharja, and Eckhardt, 2021). However, despite the significance of the earliest human civilizations and the earliest writings of ‘cuneiform’ inscribed in clay tablets, Mesopotamian history and the small cuneiform artefacts are less familiar and perhaps less visually appealing to many museum visitors, and there are almost no Mesopotamian games or virtual experiences (Rhodes and Woolley, 2023).

\section{The Future Museum}
We speculate that museums of the not-too-distant future may support easy access to digital ‘takeaways’ of individual artefacts. For example, as shown in Figure 2, QR codes to download artefacts of interest.

\begin{figure}[h]
    \centering
    \includegraphics[width=1\linewidth]{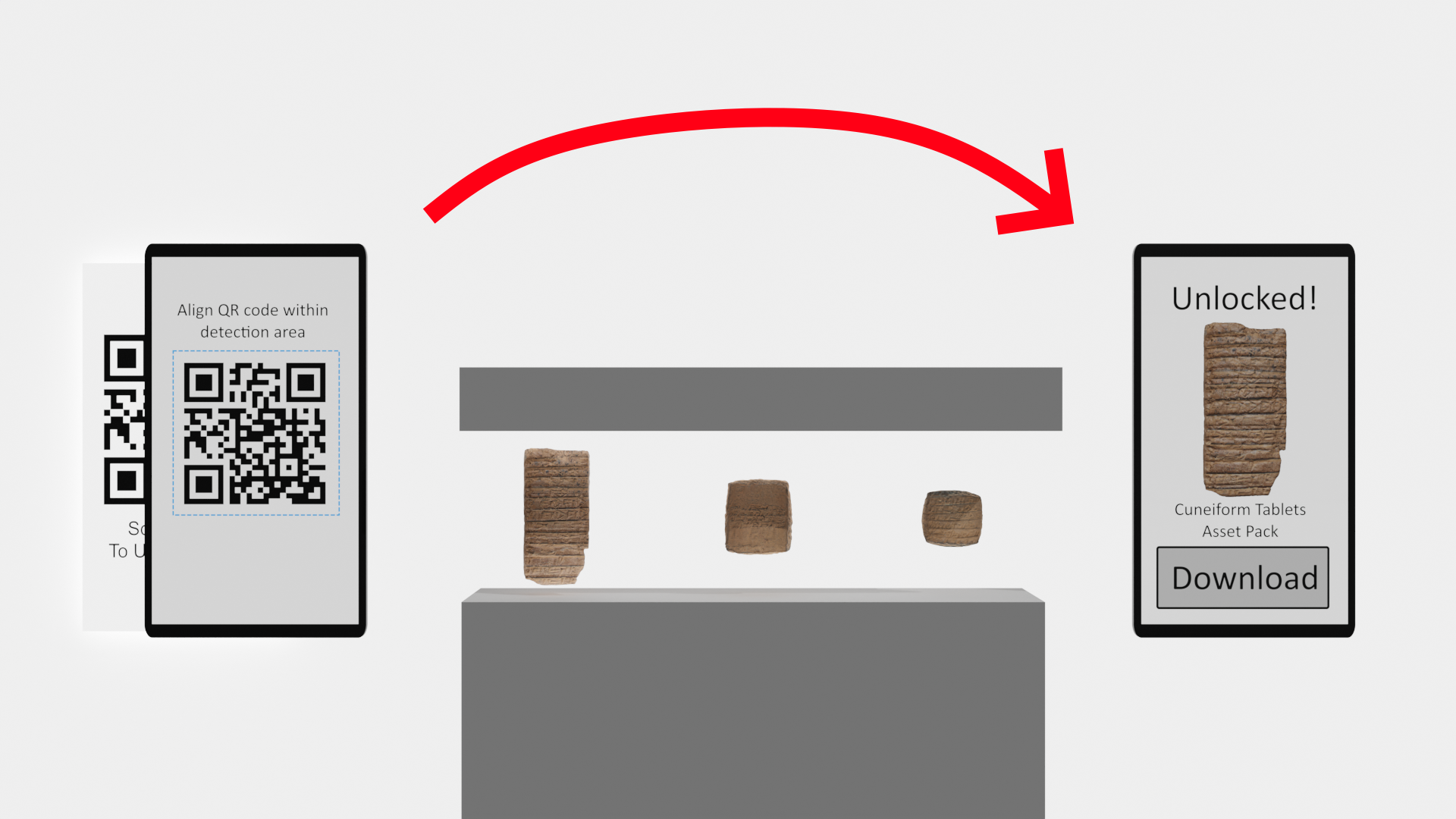}
    \caption{‘Takeaway’ 3D asset pack of cuneiform artefacts can be unlocked by scanning QR codes. Artefact scans are courtesy of National Museums Liverpool (World Museum).}
\end{figure}

Our vision includes visitor’s ability to 'collect', that is to unlock digital representations of, a variety of \textbf{Artefacts} from many time periods. In addition, these models will be supplemented with other thematically appropriate items, such as \textbf{Curation Assets} and \textbf{Experiences}. The models of the artefacts can be high-quality 3D scans, which accurately represent the real-life artefacts.

\textbf{Artefacts} are generally 3D models of real-world physical tangible heritage. They are the result of some form of object acquisition process, such as laser scanning or photogrammetry.

\textbf{Curation Assets} (referred hereafter to as 'Assets') are items allowing the user to enhance their own virtual museum. These may include appropriately themed set dressings, such as display stands, wall and floor textures, lighting, and other static items that are not artefacts, i.e., a suit of medieval armour.

\textbf{Experiences} are other content relating to, and with the focus on, the artefacts and their history. Experiences may include images or videos but will ideally be some form of interactive or immersive encounter. 

A museum visit could lead to downloadable options of recommended media and experiences, or entire ‘asset packs’ of themed artefacts, related experiences and even virtual décor to create a themed gallery. For example, one Mesopotamian-themed asset pack could contain some artefacts, themed assets such as reed baskets and pottery, and an experience allowing you to play The Royal Game of Ur. This is an Ancient Mesopotamian board game, the rules to which were decoded from a cuneiform tablet.

The assets that are collected along with the artefacts, could enhance a visit while in the virtual museum. These assets could be as simple as the information on the small card next to the physical artefact, to a video on how the artefact was created or used. This 'asset pack' will allow visitors to curate their own virtual museum, with their own virtual exhibitions. An example of an asset pack, designed to improve engagement in Mesopotamian history is shown in Figure 3. We aim for more enthusiasm, especially for those smaller and often overlooked, but no less interesting, items. The simpler assets can be easily achieved, however, more complex assets would need to be planned and created, and the extent of these may be curtailed by funding.

\begin{figure}[h]
    \centering
    \includegraphics[width=1\linewidth]{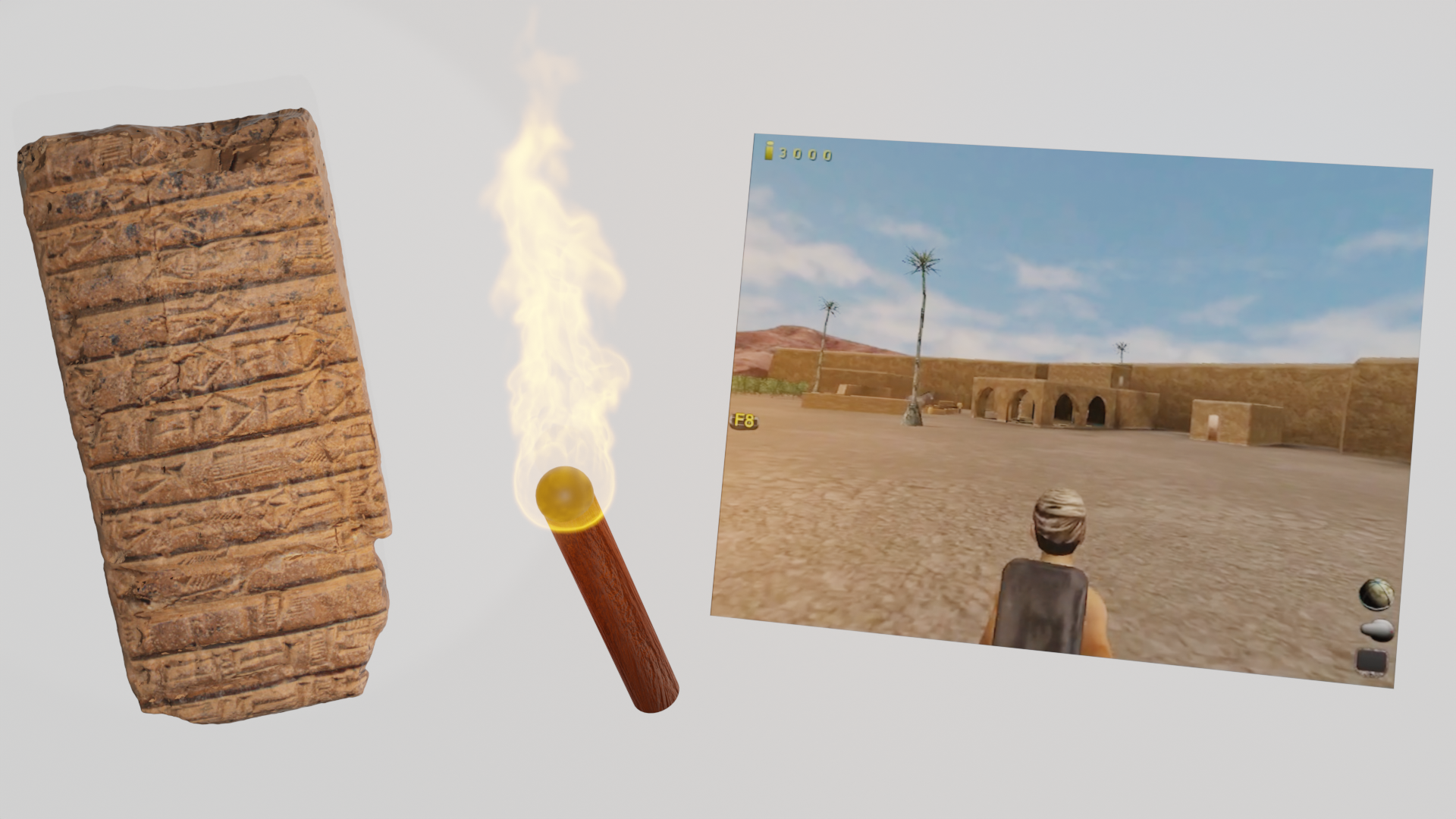}
    \caption{Asset pack content examples - Cuneiform tablet \textbf{artefact} (left), gallery lighting \textbf{curation asset} (middle), and Mesopotamian game \textbf{experience} (right).}
\end{figure}

As illustrated in Figure 4., virtual galleries can interconnect with each other and with portals to experiences, media, and information. For example, as shown in the figure, a modern-style gallery connects to an Egyptian experience and interconnects with a torch-lit historical gallery with a portal to a Mesopotamian experience.

\begin{figure}[h]
    \centering
    \includegraphics[width=1\linewidth]{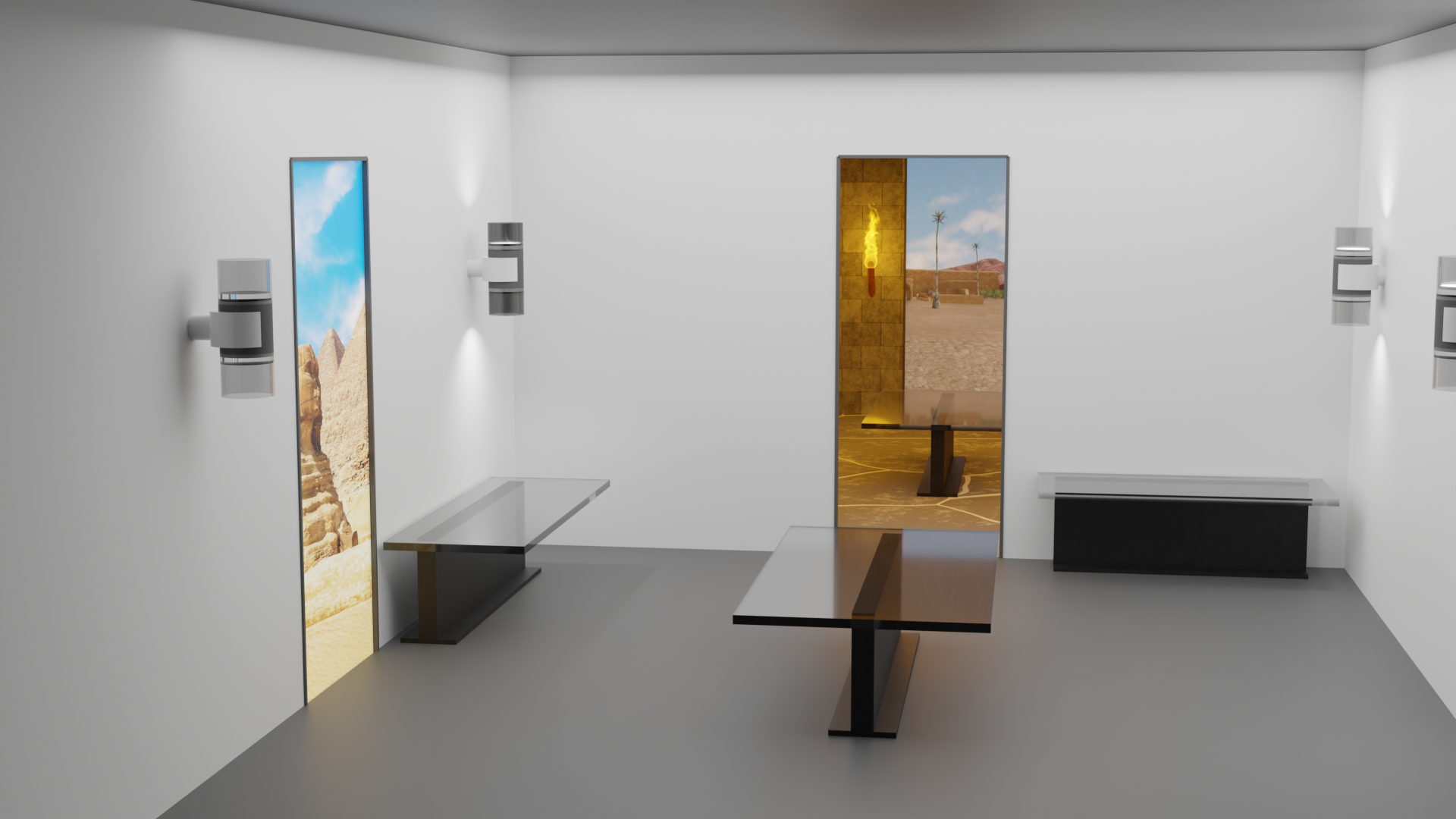}
    \caption{Virtual galleries interconnect with each other and with ‘experience portals’ }
\end{figure}

In the next section we speculate on this future virtual museum, starting with a visit to a traditional physical museum. 

\subsection{A Visit to the Future Museum}
Alice visits a museum. She walks through the Egyptian and Near Eastern exhibits, where ancient artefacts line the walls. Some artefacts about ancient writing catch her interest, and she spends some time learning about them. Hanging on the wall nearby is a poster with a QR code. Alice takes out her phone and scans it. A message pops up on the screen informing her that the museum has an asset pack for ancient writing, which contains artefact models, media, and experiences. The downloaded models can be kept and used in Alice’s own personal virtual museum. 
 
After meandering through the museum and collecting more artefacts, Alice heads to the museum’s café. Over a coffee, she pulls out her phone, and browses the new artefacts she has downloaded. One appears especially interesting, and she’d like to see it in more detail. Using AR she can project it out onto the table in front of her (Figure 5). 

In the AR view, Alice is able to see some of the finer detail, extending her phone forward to zoom in on the model, a lot more clearly than she could by looking at the real one behind a pane of glass.

\begin{figure}[h]
    \centering
    \includegraphics[width=1\linewidth]{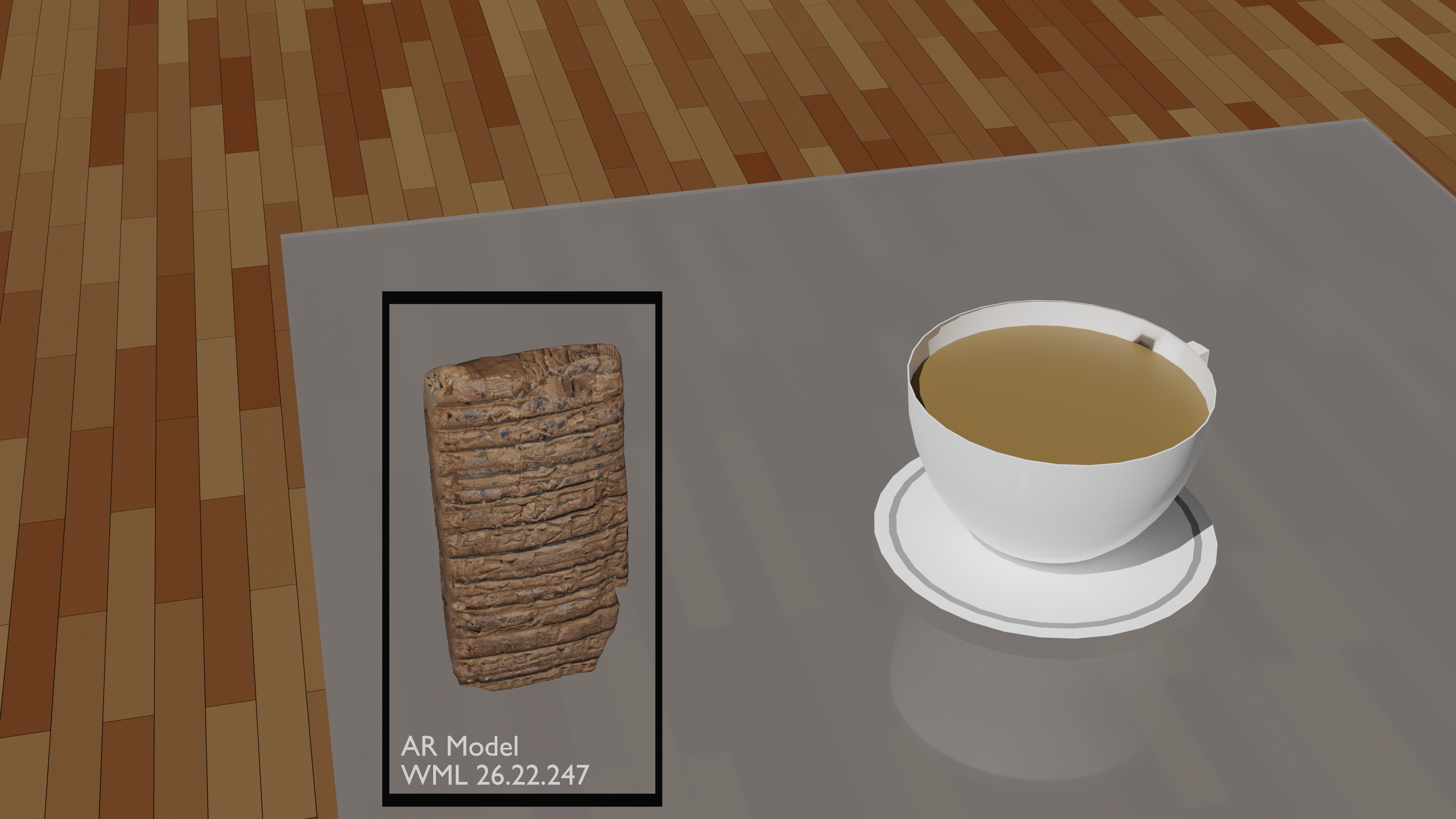}
    \caption{Over coffee, cuneiform tablet viewed with AR}
\end{figure}

Later, Alice arrives at home, turns on her PC, and loads the Personal Museum Application. She spends some time curating her own exhibit from the models she collected today. She also browses through her collection of models, and decides to include other artefacts that fit the theme. She uses some of the assets that came with the artefacts to get everything exactly how she wants it to look. Alice knows her friend from another country, Bob, is also interested in this period of history, so she shares it with him.  

Bob appreciates Alice sending her exhibition and, in return, sends his own exhibition, containing similar artefacts from his local museum. This gives Alice another museum to add to her list of museums to visit in person. Loading Bob’s exhibition, Alice puts on her VR headset, and walks around the virtual museum, comparing it to her earlier experience of walking around the physical museum. 

\section{Implications of Virtual Citizen Curated Future Museums}

\subsection{Future Museum Benefits}
\textbf{Accessibility.} One benefit is the improvement in accessibility. Those who are visually impaired may enlarge the artefacts for better viewing. Those with physical disabilities may have difficulty moving around a physical museum, so a personal virtual museum may be a more appealing option. 

\textbf{Engagement.} Virtual museums can draw attention to otherwise overlooked artefacts. As with the case study, cuneiform tablets which are historically significant yet often neglected can receive greater visibility through this system.

\textbf{Interactivity.} Digital space expands more easily than physical galleries, which face limitations that cannot currently meet the increasing cultural needs of the public, as noted by Wang and Zhu (2022). As Woolley et al. (2021) observe, it is typically the case that 95\% of a museum’s collection is not on display. Digital copes allow all artefacts to be shown, offering unparalleled versatility and accessibility (Figure 6).

\begin{figure}[h]
    \centering
    \includegraphics[width=1\linewidth]{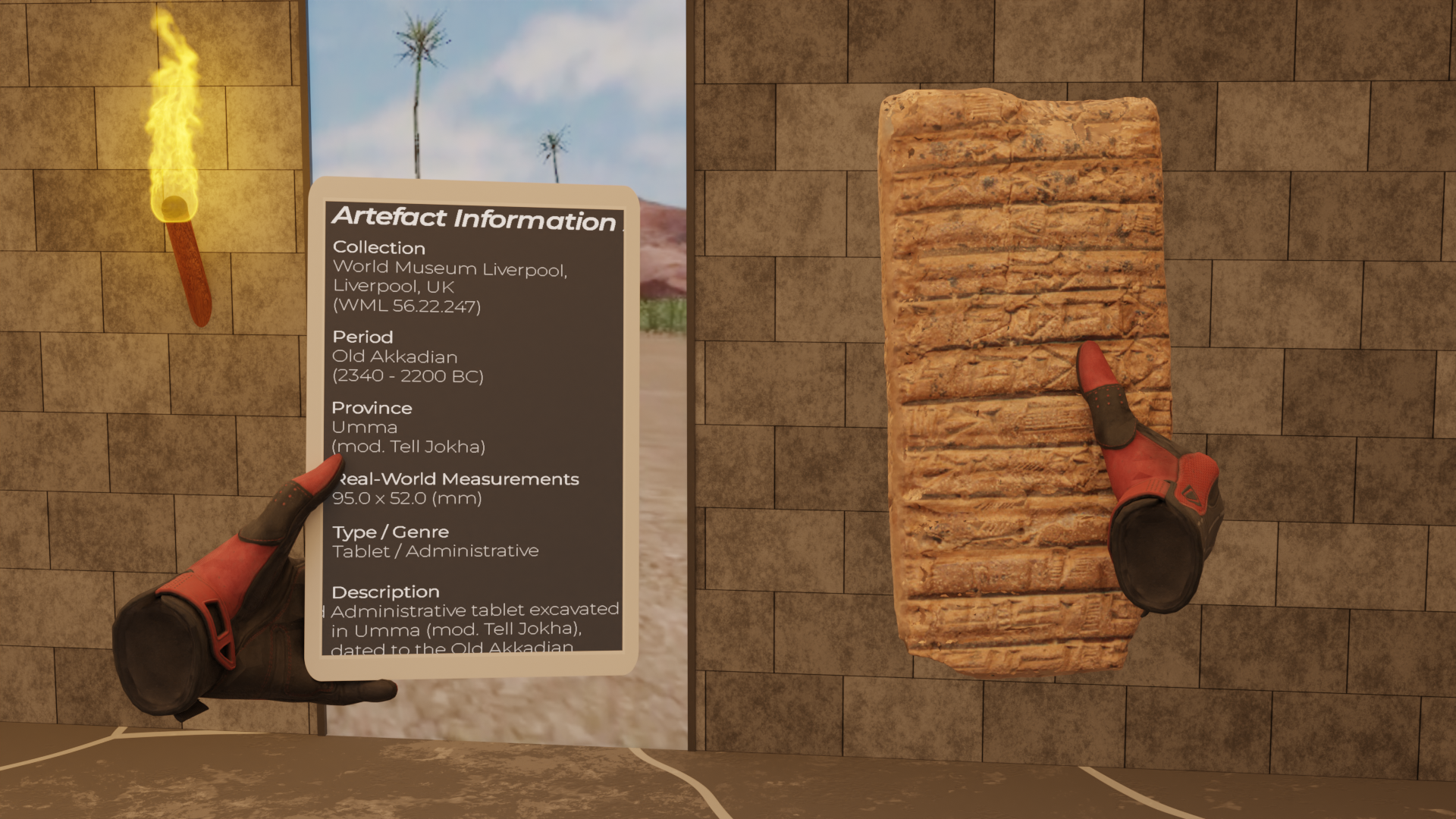}
    \caption{Artefacts are not locked in display cases in the Virtual Museum, and can be handled directly.}
\end{figure}

\textbf{Intangible Heritage Integration.} Another benefit is the opportunity to integrate intangible heritage. Intangible heritage refers to the cultural practices of people, for example, the ancient methods of writing cuneiform into tablets, of weaving baskets of reeds, and the like are all practices that may have survived in one form or another, but the experiences provided in asset packs could include these ancient traditions and crafts.

\subsection{Future Museum Challenges}
\textbf{Digitising Artefacts.} The digitisation of artefacts, particularly the creation of high-quality, all-around 3D models, is a significant challenge. In a sector with limited resources, the time-investment required for 3D model creation and the high-costs of good quality 3D scanners, pose significant challenges. One potential solution for this problem is the development of low-cost 3D model acquisition systems, such as those presented by Collins, et. al. (2019). 

\textbf{Investment.} Another major challenge is the lack of funding for projects such as this one. Digital Heritage is underfunded, so the means to create and deploy this system may be limited. This also contributes to many other issues, such as hosting the system and employing people for various tasks. 

\textbf{Curation of Asset Packs.} Some asset pack components may be readily achievable; however, others could require substantial digitization effort or design and development. Alongside assets that need creating, experiences would also need to be either created or adapted for such a system. For example, the Discover Babylon game was created in 2004, so difficulties would need to be overcome to update it for a modern audience. These difficulties increase when porting experiences to other mediums, such as VR.

\textbf{Impact on Traditional Museums.} We must also ask ourselves about how traditional museums may change, however, it is not our desire to replace traditional with virtual. We could speculate that recommendation systems within future virtual museums might entice users to visit physical museums which previously they may not have considered. Similarly, museums could attract visitors by offering access to new asset pack content, which visitors would need to visit the physical museum to collect. In this way we envisage that traditional museums would remain fundamental to the enablement of interest and engagement in artefacts and their histories. 

\section{Discussion and Future Research}
Across sectors, from education and media to retail, digital innovation brings myriad opportunities and challenges, and museums are no exception. Perhaps one of the commonalities is that future in-person visits and encounters will likely focus on meaningful and shared experiences. Future planned research involves the elicitation of stakeholder opinions on future heritage experiences and future virtual museums, including citizen-curated museums. 

\section{ACKNOWLEDGEMENTS}
The authors thank National Museums Liverpool (World Museum) for permission to use the artefact models. For the purposes of open access, the authors have applied a Creative Commons Attribution (CC-BY) licence to any Accepted Author Manuscript version arising from this submission.

\end{document}